\documentclass[twocolumn,prl,aps,showpacs
  ,superscriptaddress,nofootinbib]{revtex4}

\usepackage{amsmath}
\usepackage{graphicx}
\usepackage{hyperref}
\usepackage{eimodels}
\usepackage{eivars}
\usepackage{xspace}

\hypersetup{
    bookmarks=true,         
    unicode=false,          
    pdftoolbar=true,        
    pdfmenubar=true,        
    pdffitwindow=false,     
    pdfstartview={FitH},    
    pdfnewwindow=true,      
    colorlinks=true,        
    linkcolor=red,          
    citecolor=cyan,         
    filecolor=magenta,      
    urlcolor=blue,         
    linktocpage=true
}

\newcommand{\ie}{i.e.\xspace}

\newlength{\wsingfig}
\newlength{\wdblefig}
\newcommand{\boldmathsymbol}[1]{{\ensuremath{\boldsymbol{#1}}}}

\newcommand{\sss}[1]{{\scriptscriptstyle{#1}}}

\newcommand{\EI}{\textit{Encyclop{\ae}dia Inflationaris}\xspace}
\newcommand{\CAMB}{\texttt{CAMB}\xspace}
\newcommand{\COSMOMC}{\texttt{COSMOMC}\xspace}
\newcommand{\ASPIC}{\texttt{ASPIC}\xspace}
\newcommand{\MULTINEST}{\texttt{MultiNest}\xspace}

\newcommand{\MeV}{\mathrm{MeV}}
\newcommand{\GeV}{\mathrm{GeV}}

\newcommand{\TeV}{\mathrm{TeV}}

\newcommand{\Mp}{M_\usssPl}

\newcommand{\ubest}{\sss{\mathrm{best}}}

\newcommand{\calP}{\mathcal{P}}
\newcommand{\calM}{\mathcal{M}}
\newcommand{\calE}{\mathcal{E}}
\newcommand{\calEbest}{\mathcal{E}_{\ubest}}

\newcommand{\ucos}{\mathrm{cos}}

\newcommand{\tetas}{\boldmathsymbol{\theta}_{\ucos}}
\newcommand{\tetai}{\boldmathsymbol{\theta}_{\uinf}}
\newcommand{\tetar}{\boldmathsymbol{\theta}_{\ureh}}

\begin{document}

\title{Observing the Inflationary Reheating}

\author{J\'er\^ome Martin}
\email{jmartin@iap.fr}
\affiliation{Institut d'Astrophysique de Paris, UMR 7095-CNRS,
Universit\'e Pierre et Marie Curie, 98 bis boulevard Arago, 75014
Paris, France}

\author{Christophe Ringeval}
\email{christophe.ringeval@uclouvain.be}
\affiliation{Centre for Cosmology, Particle Physics and Phenomenology,
  Institute of Mathematics and Physics, Louvain University, 2 Chemin
  du Cyclotron, 1348 Louvain-la-Neuve, Belgium}

\author{Vincent Vennin}
\email{vincent.vennin@port.ac.uk}
\affiliation{Institute
  of Cosmology \& Gravitation, University of Portsmouth, Dennis Sciama
  Building, Burnaby Road, Portsmouth, PO1 3FX, United Kingdom}

\date{\today}

\begin{abstract}
  Reheating is the the epoch which connects inflation to the
  subsequent hot Big-Bang phase. Conceptually very important, this era
  is, however, observationally poorly known. We show that the current
  Planck satellite measurements of the Cosmic Microwave Background
  (CMB) anisotropies constrain the kinematic properties of the
  reheating era for most of the inflationary models. This result is
  obtained by deriving the marginalized posterior distributions of the
  reheating parameter for about $200$ models taken in {\EI}. Weighted
  by the statistical evidence of each model to explain the data, we
  show that the Planck 2013 measurements induce an average reduction
  of the posterior-to-prior volume by $40\%$. Making some additional
  assumptions on reheating, such as specifying a mean equation of
  state parameter, or focusing the analysis on peculiar scenarios, can
  enhance or reduce this constraint. Our study also indicates that the
  Bayesian evidence of a model can substantially be affected by the
  reheating properties. The precision of the current CMB data is
  therefore such that estimating the observational performance of a
  model now requires incorporating information about its reheating
  history.
\end{abstract}

\maketitle

The recent release of high accuracy Cosmic Microwave Background (CMB)
data by the Planck satellite~\cite{Ade:2013zuv} has made it possible
to drastically improve our knowledge of inflation, in particular the
slow-roll phase during which the expansion of the early Universe is
accelerated~\cite{Ade:2013uln, Martin:2013nzq}. But how inflation ends
remains observationally poorly known. The so-called reheating
era~\cite{Turner:1983he, Traschen:1990sw, Kofman:1997yn,Bassett:2005xm,
  Braden:2010wd, Frolov:2010sz, Allahverdi:2010xz, Drewes:2013iaa, Amin:2014eta, Hertzberg:2014iza} is
conceptually of major significance for several reasons. Reheating
explains how inflation is connected to the subsequent hot Big-Bang
phase and drives the production of all types of matter at its
onset. Because the micro-physics of reheating depends on the
interaction between the inflaton and the other fundamental fields, by
constraining this era, one can learn about these
couplings. Furthermore, reheating is sensitive to the shape of the
inflationary potential in a field regime that is different from where
slow-roll inflation takes place. Finally, once the inflaton decay
products have thermalized, the radiation dominated era starts and, for
the first time in its history, the Universe as a whole acquires a
temperature. Measuring this ``reheating temperature'' is of crucial
importance to understanding the thermal history of the Universe.

For all these reasons, any experimental constraint on the reheating
era is highly desirable. In the present letter, we make use of the
method developed in Refs.~\cite{Martin:2006rs,
  Lorenz:2007ze,Ringeval:2007am,Martin:2010kz,Demozzi:2012wh,
  Ringeval:2013hfa} (see also Refs.~\cite{Easther:2011yq, Dai:2014jja})
and show that the Planck 2013 CMB data put non-trivial constraints on
the reheating era for essentially all the slow-roll single-field
models, which are the scenarios preferred by the
data~\cite{Martin:2013nzq, Martin:2014vha, Giannantonio:2014rva}.

Constraints on the reheating stage from CMB data have been first
discussed in Refs.~\cite{Martin:2006rs,Martin:2010kz} using the WMAP
three- and seven-year
measurements~\cite{Spergel:2006hy,Jarosik:2010iu}. It was shown that,
for the small and large-field inflationary models, reheating histories
exhibiting a negative equation of state parameter were constrained to
have a reheating temperature higher than the TeV energy scale. Since
then the situation has significantly improved, notably thanks to the
Planck $2013$ data release~\cite{Ade:2013zuv} but also to our ability
to derive reheating-consistent observational predictions for a much
wider survey of inflationary scenarios~\cite{Martin:2013nzq,
  Martin:2014vha}.

In the following, we make use of the Planck 2013 data to derive the
posterior probabilities of the reheating parameters associated with
almost $200$ inflationary models taken from the
{\EI}~\cite{Martin:2013nzq}. Such a number is representative of all
the single-field slow-roll models with canonical kinetic term that
have been proposed so far and enables us to extract new constraints and
to draw generic conclusions on the inflationary reheating within
slow roll. So far, results were known only for very peculiar reheating
histories and/or priors~\cite{Ade:2013uln} and, therefore, our work
represents the first general study of how Planck $2013$ can constrain
the end of inflation.

\par

Let us now see how the reheating phase affects inflationary
observables. Within a given inflationary model, and for fixed values
of the parameters characterizing the shape of the potential, cosmic
inflation stops at a well-determined energy density $\rhoend$. The
redshift $\zend$ at which this occurs is of crucial importance as it
relates the physical value of any length scale measured today to those
during inflation. Denoting by the index ``reh'' the end of the
reheating era, straightforward manipulations yield
\begin{equation}
\begin{aligned}
1+ \zend & = \dfrac{\areh}{\aend}(1+\zreh) =
\dfrac{\areh}{\aend} \left(\dfrac{\rhoreh}{\rhotildegamma}
\right)^{1/4} \\ & = \dfrac{\areh}{\aend}
\dfrac{\rhoreh^{1/4}}{\rhoend^{1/4}}
\dfrac{\rhoend^{1/4}}{\rhotildegamma^{1/4}} \equiv
\dfrac{1}{\Rrad} \left(\dfrac{\rhoend}{\rhotildegamma}
\right)^{1/4},
\end{aligned}
\label{eq:zend}
\end{equation}
where $a$ is the Friedmann-Lema\^{\i}tre-Robertson-Walker scale
factor. The quantity $\rhotildegamma$ stands for the energy density of
radiation today rescaled by the number of relativistic degrees of
freedom. Such an expression shows that, even within a completely
specified inflationary scenario, $\zend$ and thus all inflationary
observables, are affected by $\zreh$. The last line of
Eq.~\eqref{eq:zend} should be understood as a definition of the
reheating parameter $\Rrad$. It equals unity either for instantaneous
reheating ($\rhoreh=\rhoend$) or if reheating is radiation
dominated. As shown in Ref.~\cite{Martin:2010kz}, the reheating
parameter also verifies
\begin{equation}
\label{eq:Rrad}
\ln \Rrad=\frac{\Delta N}{4}\left(3\wrehbar - 1\right)=
\frac{1-3\wrehbar}{12\left(1+\wrehbar\right)}
\ln\left(\frac{\rhoreh}{\rhoend}\right)\, ,
\end{equation}
where $\Delta N \equiv \Nreh-\Nend$ is the duration of reheating in
e-folds ($N = \ln a$) and $\wrehbar$ is the mean value of the equation of
state parameter defined by
\begin{equation}
\wrehbar \equiv \dfrac{1}{\Delta N} \int _{\Nend}^{\Nreh} 
\dfrac{P(n)}{\rho(n)}\ud
n,
\label{eq:wrehbar}
\end{equation}
where $P$ is the total pressure. All these expressions are fully
generic and do not assume anything about the microphysics of the
reheating process. In addition, as shown in Refs.~\cite{Martin:2010hh,
  Kuroyanagi:2013ns, Kuroyanagi:2014qza}, the above parametrization
is the most generic as it remains valid even in presence of any
additional entropy production eras that could occur after reheating.

\par

Let us now explain how constraints on reheating can be
inferred. Each inflationary model $\calM_i$ is characterized by some
parameters $\tetai$, describing the slow-roll phase, and $\tetar$,
describing the reheating phase. A complete cosmological scenario also
includes the post-inflationary history, characterized by the
cosmological parameters $\tetas$. Here the $\tetas$ have been chosen to be those of
a flat $\Lambda$CDM Universe complemented by the astrophysical and
experimental nuisance parameters associated with the Planck
satellite~\cite{Ade:2013zuv}. The inflationary models considered in
our analysis are listed in Ref.~\cite{Martin:2013nzq} in which the
number and physical meaning of the $\tetai$ are detailed. As
mentioned above, the most generic parametrization of reheating is
given by only one parameter $\Rrad$. However, from a data analysis
point of view, it is more convenient to consider $\tetar=\Rreh$ where
\begin{equation}
\Rreh \equiv \Rrad \dfrac{\rhoend^{1/4}}{\Mp}
\label{eq:Rreh}
\end{equation}
is a rescaled reheating parameter~\cite{Martin:2006rs,
  Martin:2010hh}. Within each $\calM_i$, the energy at the end of
inflation is completely specified and both parameters, $\Rrad$ and
$\Rreh$, are in one-to-one correspondence. The advantage of $\Rreh$
over $\Rrad$ is that it minimizes degeneracies in parameter
space. Starting from some motivated prior probability distributions
for each $\calM_i$, the Planck CMB data, $D$, can be used to derive
the posterior probability distributions in the parameter space
$\{\tetai,\tetar,\tetas\}$. By marginalization over the $\tetai$ and
$\tetas$, one finally obtains the marginalized posterior
$\calP(\tetar|D,\calM_i)$ we are interested
in~\cite{Trotta:2005ar,Trotta:2008qt}. If this posterior is
``more peaked'' than the prior $\pi(\tetar)$ then the data provide us
with some non-trivial information on reheating.

In practice, the prior distributions for the $\tetai$ have been chosen
exactly as in Ref.~\cite{Martin:2013nzq} while the prior for the
cosmological, astrophysical and experimental nuisance parameters are
the same as in Ref.~\cite{Ade:2013zuv}. The prior on $\Rreh$ follows
from the requirements that $\rhonuc<\rhoreh<\rhoend$, where
$\rhonuc\simeq (10\, \MeV)^{4}$ and $-1/3<\wrehbar<1$. From
Eqs.~\eqref{eq:Rrad} and \eqref{eq:Rreh}, this leads to
\begin{equation}
\ln \left(\frac{\rhonuc^{1/4}}{\Mp}\right)
<\ln \Rreh <-\frac13 \ln \left(\frac{\rhonuc^{1/4}}{\Mp}\right)
+\frac43 \ln \left(\frac{\rhoend^{1/4}}{\Mp}\right).
\end{equation}
Since the order of magnitude of $\Rreh$ is a priori unknown, we have
chosen a uniform prior on $\ln \Rreh$ in the above range. Concerning
data analysis, we have used the public likelihood provided by the
Planck Collaboration~\cite{Ade:2013kta}. In order to perform $200$
data analyses of the Planck data, one for each $\calM_i$, we have
followed the method detailed in Ref.~\cite{Ringeval:2013lea}. It
requires the evaluation of a marginalized likelihood in the slow-roll
parameter space followed by nested sampling analysis for each model
$\calM_i$. For this purpose, we have used modified versions of the
{\CAMB}~\cite{Lewis:1999bs}, {\COSMOMC}~\cite{Lewis:2002ah} and
{\MULTINEST}~\cite{Feroz:2008xx} codes as well as our public library
{\ASPIC}~\cite{Martin:2014vha}.

\begin{figure}
\begin{center}
\includegraphics[width=\wdblefig]{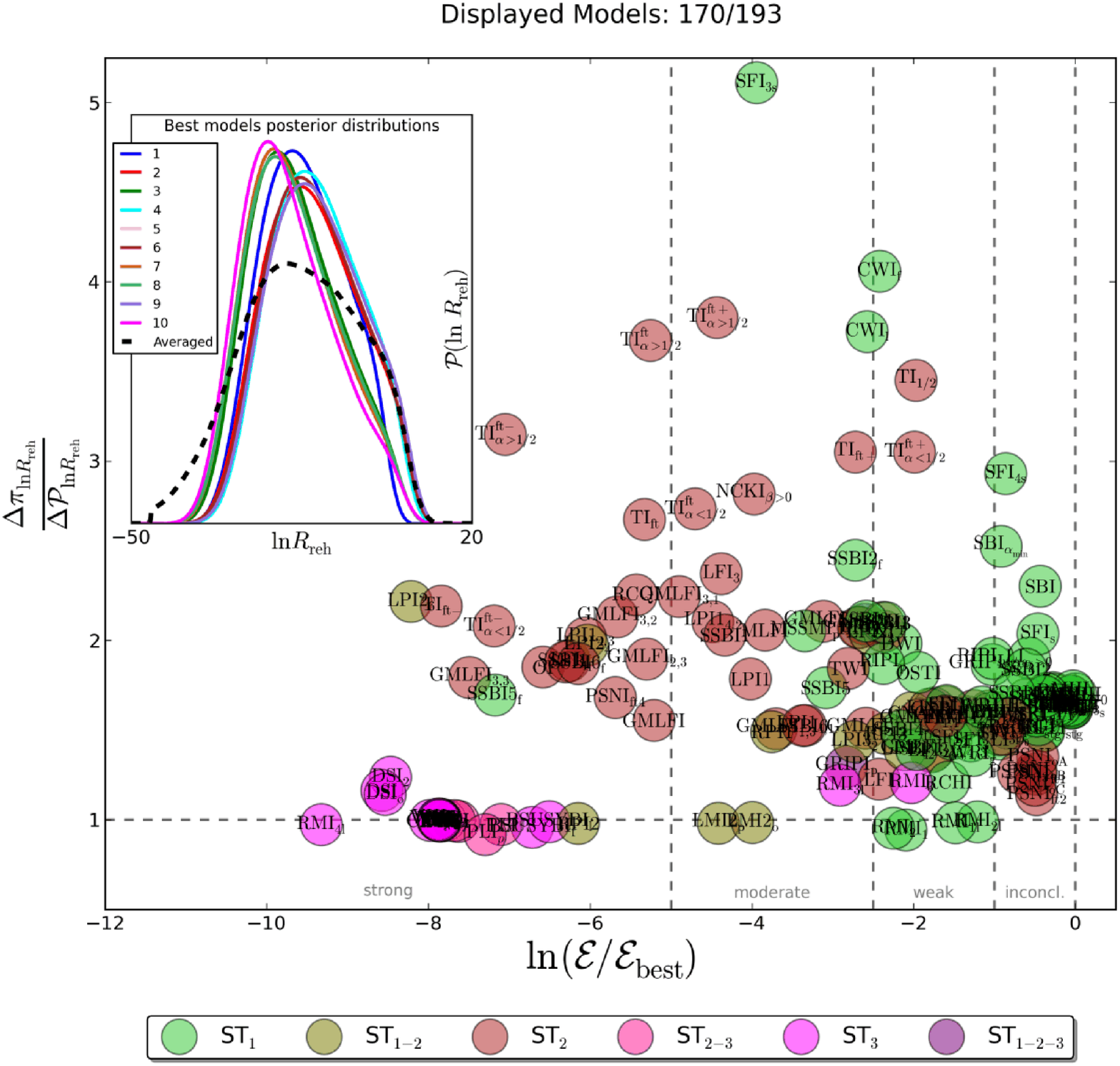}
\caption{Prior-to-posterior width ratio of the reheating parameter
  $\ln \Rreh$ versus the logarithm of the Bayesian evidence for the
  {\EI} scenarios. Each model is represented by a circle, the color of
  which refers to the Schwarz-Terrero Escalante
  classification~\cite{Schwarz:2001vv}, and an acronym matching the
  {\EI} classification~\cite{Martin:2014vha}. The vertical dashed
  lines separate the four Jeffreys' categories: inconclusive, weakly
  disfavored, moderately disfavored and strongly disfavored, from
  right to left.  The dashed horizontal line corresponds to a
  prior-to-posterior width ratio equal to unity. Models above this
  line have a reheating stage which is constrained (the higher in the
  plot the more it is constrained). The inset displays the posterior
  distributions of $\ln\Rreh$ for the ten best Planck $2013$ models
  ($\kmiii$, $\esisqrtTWO$, $\biSIXs$, $\mhis$, $\bis$, $\esi$, $\biFIVEs$, $\kkltis$, $\kmii$, $\biFOURs$). The thick black dashed line corresponds to
  the averaged distribution over all {\EI} models weighted by their
  Bayesian evidence.}
\label{fig:evidVScp}
\end{center}
\end{figure}

\par

Let us now turn to the results. In Fig.~\ref{fig:evidVScp}, we have
represented the prior-to-posterior standard deviation ratio of $\ln
\Rreh$, $\Delta \pi_{\ln \Rreh}/\Delta \calP_{\ln \Rreh}$, versus the
logarithm of the statistical evidence, $\ln (\calE/\calEbest)$, for
all the \EI scenarios. The quantity $\Delta \pi_{\ln\Rreh}/\Delta
\calP_{\ln\Rreh}$ measures how much reheating is constrained for a
given model. Clearly, if it equals unity (see the dashed horizontal
line in Fig.~\ref{fig:evidVScp}), then the posterior is as wide as the
prior and there is no information gain. If, on the contrary, $\Delta
\pi_{\ln\Rreh}/\Delta \calP_{\ln\Rreh}>1$, then the posterior is more
peaked than the prior and the data carry information on reheating. The
quantity $\ln(\calE/\calEbest)$ describes the performance of a model
in explaining the data so that models on the right in
Fig.~\ref{fig:evidVScp} are more probable than those on the left. The
four vertical dashed lines refer to the four Jeffreys' categories
which measure strength of belief~\cite{Gordon:2007xm}. From right to
left, they correspond to models which are inconclusive, weakly
disfavored, moderately disfavored and strongly disfavored. In order to
quantify to which extent the reheating stage is constrained, we
introduce the following measure
\begin{equation}
\label{eq:defratio}
\left \langle \frac{\Delta \pi_{\ln \Rreh}}{\Delta \calP_{\ln \Rreh}}
\right \rangle\equiv 
\frac{1}{\sum_j\calE_j}
\displaystyle \sum_i
\calE_i
\left(\frac{\Delta \pi_{\ln \Rreh}}{\Delta \calP_{\ln \Rreh}}
\right)_i,
\end{equation}
which is the mean value of $\Delta \pi_{\ln \Rreh}/\Delta
\calP_{\ln\Rreh}$ weighted by the Bayesian evidence, {\ie} the mean
value in the space of models. This is a fair estimate since
inefficient models will not contribute a lot to this quantity due to
their small evidence. Numerically, one obtains $\left \langle \Delta
  \pi_{\ln\Rreh}/\Delta \calP_{\ln\Rreh} \right \rangle \simeq 1.66$
which, therefore, indicates that reheating is indeed constrained by
Planck 2013. On average, the posterior distribution of the reheating
parameter is $0.60$ times smaller than the prior corresponding to a
reduction of the prior volume by $40\%$. This is our main result.

One can also discuss how reheating is constrained within each of the
Jeffreys' categories. We find that the mean value of $\Delta
\pi_{\ln\Rreh}/\Delta \calP_{\ln\Rreh}$ is $1.65$ for the inconclusive
models, $1.63$ for the weakly disfavored models, $2.10$ for the
moderately disfavored models and $1.92$ for the strongly disfavored
models. The tendency to have stronger constraints for disfavored
models is expected. There is indeed relatively small evidence for
these scenarios because, in part, some reheating histories are in
contradiction with the data and hence are constrained.  We also see
that the result $\left \langle \Delta \pi_{\ln\Rreh}/\Delta
\calP_{\ln\Rreh} \right \rangle \simeq 1.66$ is dominated by the
inconclusive models precisely because the other models are penalized
by their small evidence.

\begin{figure*}
\begin{center}
\includegraphics[width=\wdblefig]{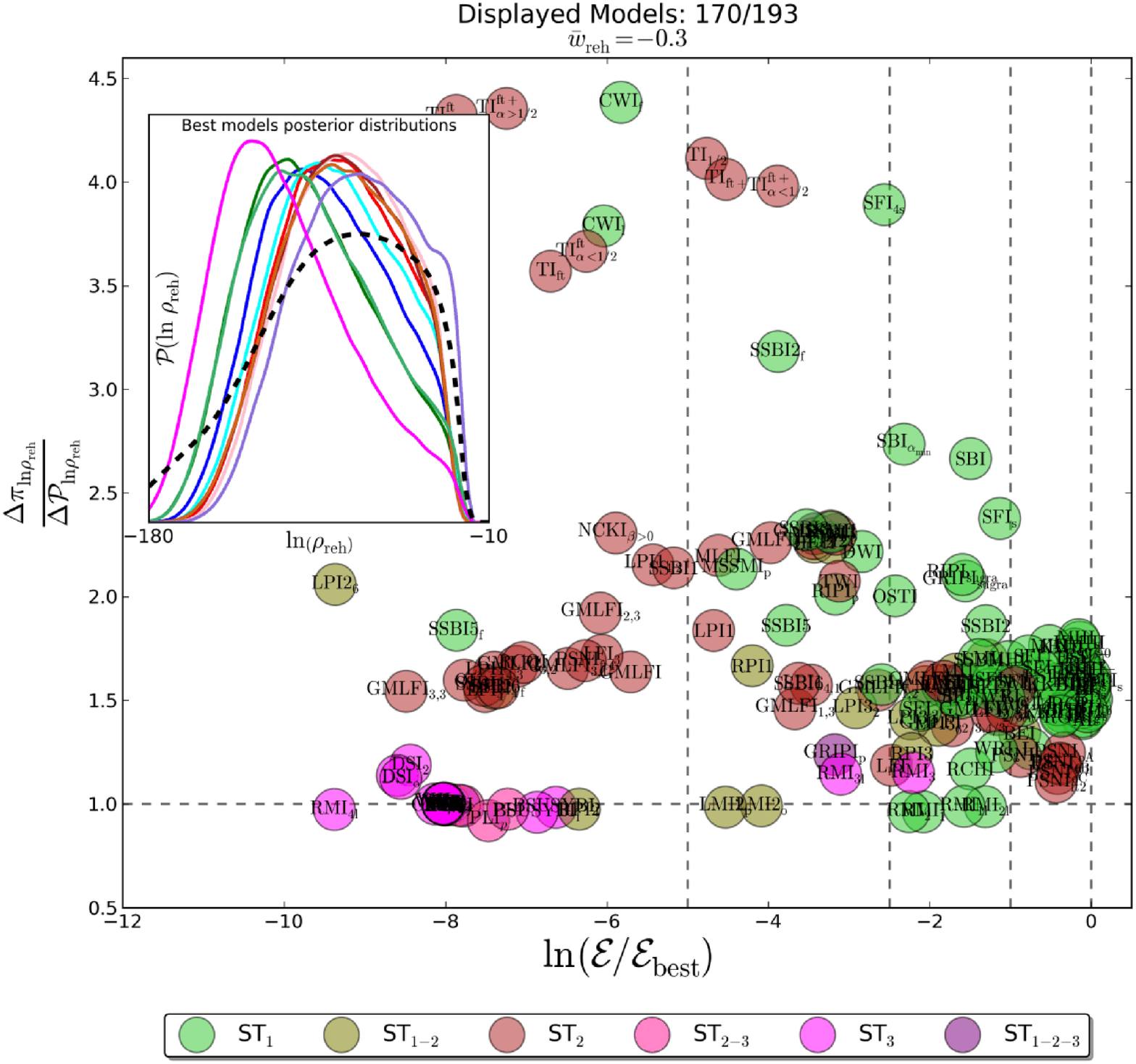}
\includegraphics[width=\wdblefig]{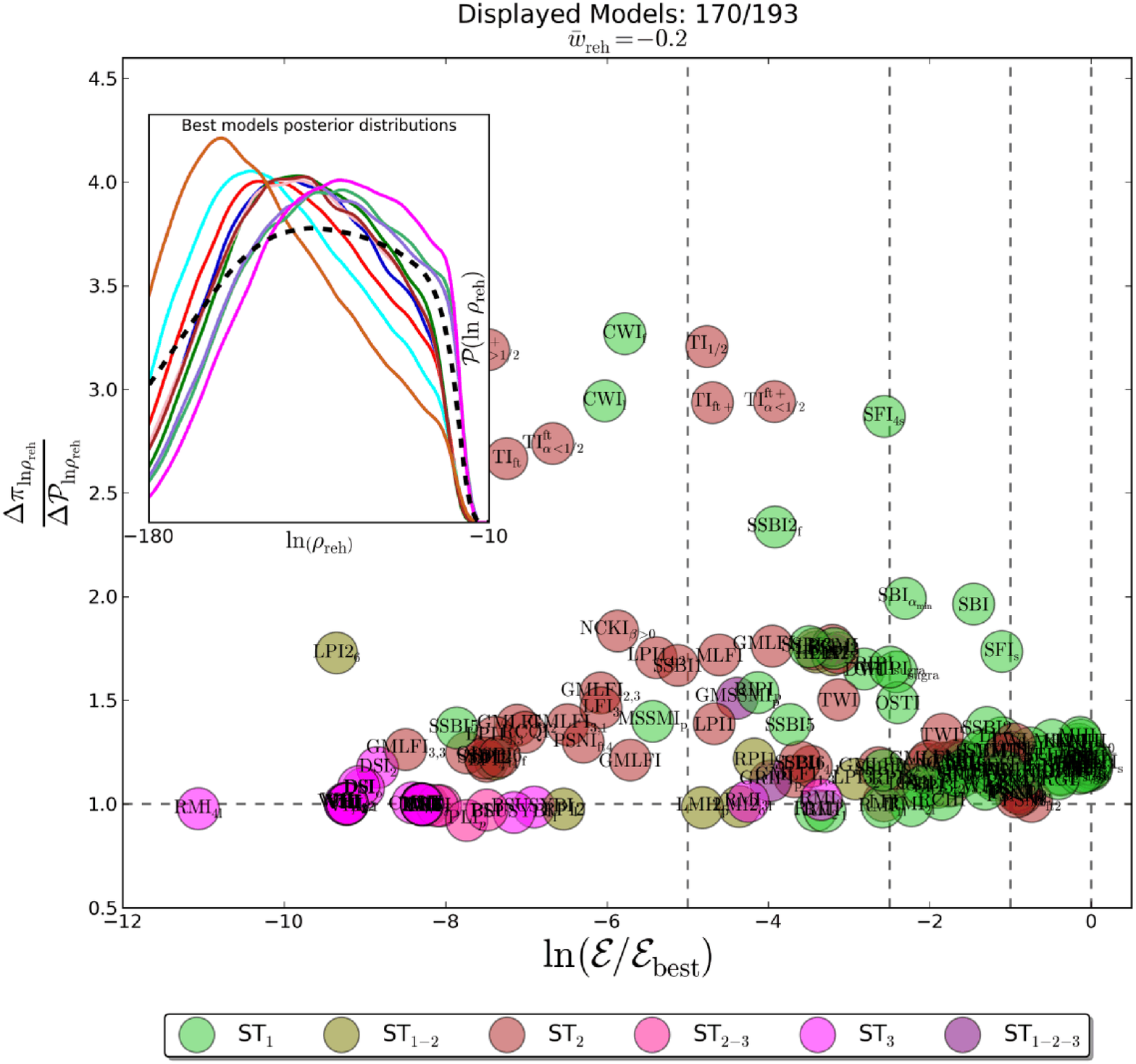}
\includegraphics[width=\wdblefig]{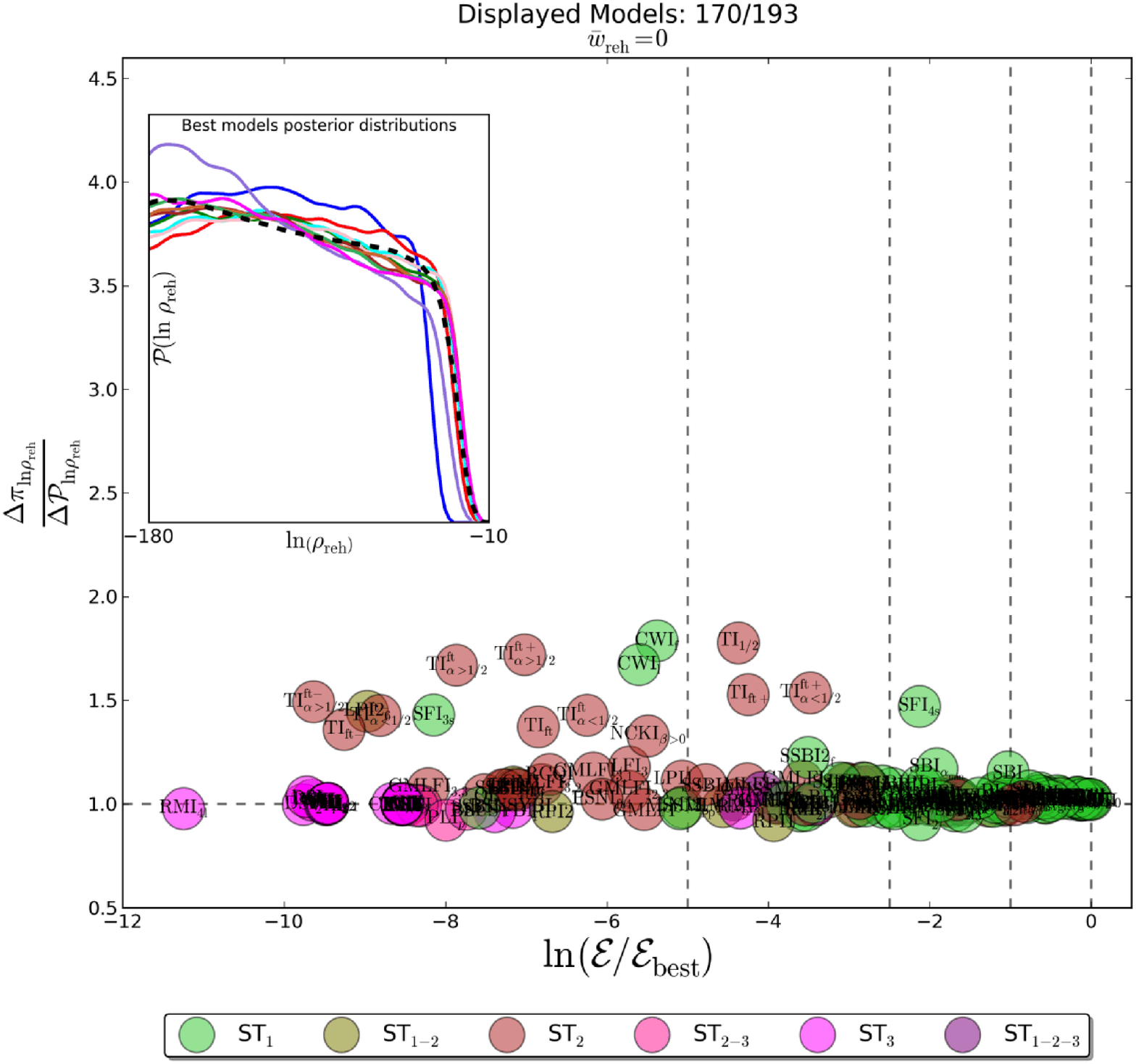}
\includegraphics[width=\wdblefig]{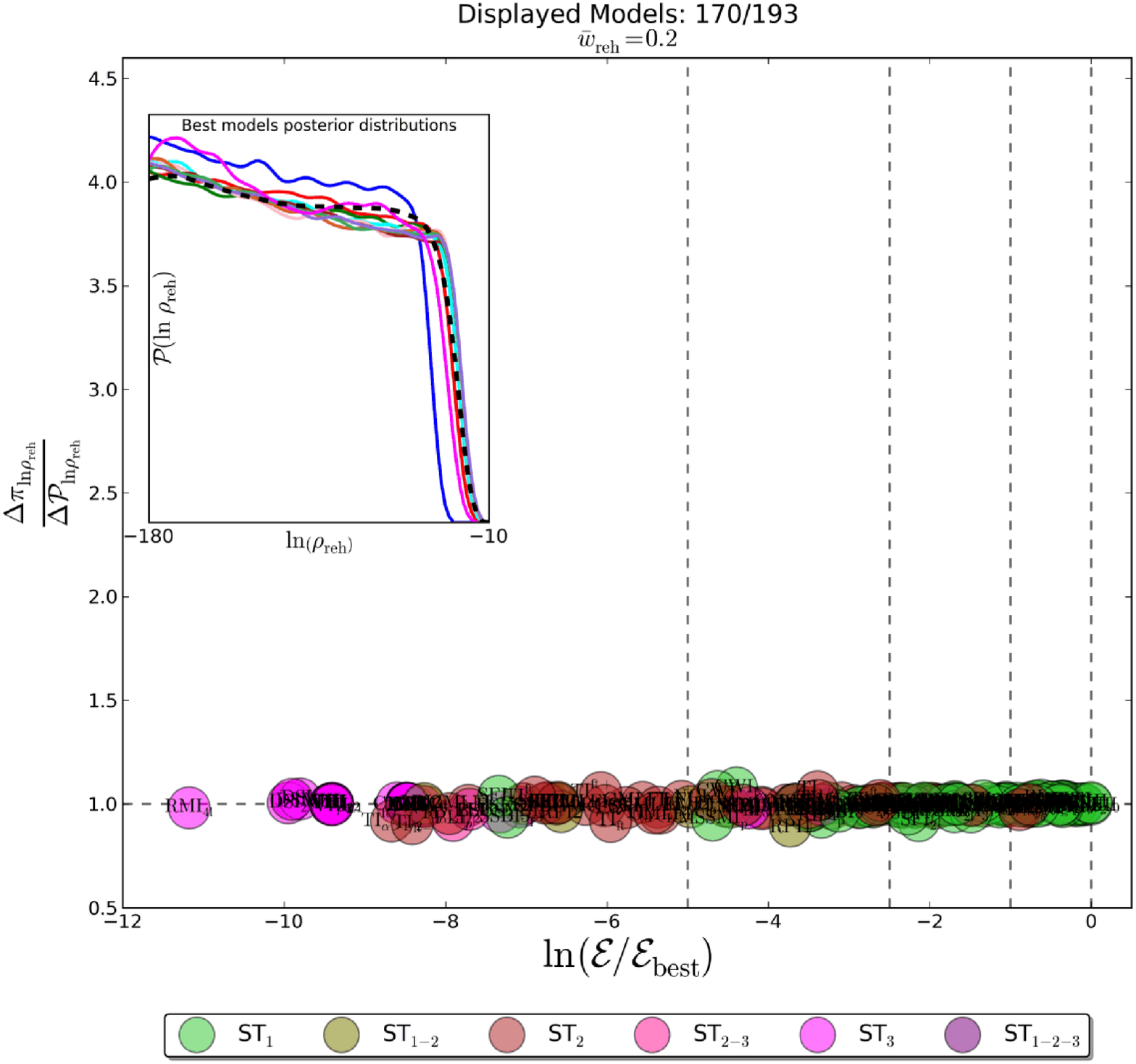}
\caption{Same as in Fig.~\ref{fig:evidVScp} but assuming the mean
  equation of state during reheating is known. The prior-to-posterior
  width for the reheating energy density $\ln(\rhoreh/\Mp^4)$ is
  represented assuming four values of the mean equation of state
  $\wrehbar$, namely $\wrehbar=-0.3$ (top left panel), $\wrehbar=-0.2$
  (top right panel), $\wrehbar=0$ (bottom left panel) and
  $\wrehbar=0.2$ (bottom right panel). The insets display the posterior
  distributions of $\ln(\rhoreh/\Mp^4)$ for the ten best models in each case, namely
  $\biTHREEs$, $\biFIVEs$, $\biTWOs$, $\biFOURs$, $\biSIXs$, $\kkltis$. $\bis$,
  $\rgis$, $\esisqrtTWOTHREE$, $\biONEs$ for $\wrehbar=-0.3$, $\biFIVEs$, $\biFOURs$,
  $\biSIXs$, $\biTHREEs$, $\bis$, $\kkltis$, $\biTWOs$, $\esio$, $\esisqrtTWOTHREE$,
  $\esisqrtTWO$ for $\wrehbar=-0.2$, $\kmiii$, $\mhis$, $\esisqrtTWO$, $\esi$, 
  $\kmiivp$, $\hi$, $\kmii$, $\esio$, $\biSIXs$, $\esisqrtTWOTHREE$ for 
  $\wrehbar=0$ and $\kmiii$, $\mhis$, $\kmiivp$, $\esi$, $\esisqrtTWO$, $\kmii$,
  $\hi$, $\esio$, $\esisqrtTWOTHREE$, $\biSIXs$ for $\wrehbar=0.3$.}
\label{fig:wfix}
\end{center}
\end{figure*}

Instead of taking the most generic parametrization, namely
$\tetar=\Rreh$, we have also performed the same analysis using a more
restrictive reheating assumption, namely that the mean equation of
state parameter $\wrehbar$ is known. In that situation, reheating
is completely specified by $\tetar=\rhoreh$, where $\rhoreh^{1/4}$
measures the reheating temperature. In Fig.~\ref{fig:wfix}, we have
represented the prior-to-posterior width ratio
$\Delta\pi_{\ln\rhoreh}/\Delta\calP_{\ln\rhoreh}$ versus the logarithm
of the evidence, $\ln(\calE/\calEbest)$ for different equation of
state parameters $\wrehbar=-0.3$, $-0.2$, $0$ and $0.2$. One obtains
$\left \langle \Delta \pi_{\ln\rhoreh}/ \Delta\calP_{\ln\rhoreh}\right
\rangle \simeq 1.55$, $1.22$, $1.03$, and $1.00$ for $\wrehbar=-0.3$,
$-0.2$, $0$ and $0.2$, respectively. Such a trend can be seen in
Fig.~\ref{fig:wfix} in which the models have a tendency to cluster
around the horizontal line $\Delta \pi_{\ln\rhoreh}/\Delta \calP_{\ln
  \rhoreh}=1$ as $\wrehbar$ increases. This means that reheating
is relatively well-constrained for $\wrehbar \le 0$ but not when
$\wrehbar$ becomes positive and approaches $1/3$. Notice that this is
expected as $\wrehbar=1/3$ corresponds to radiation-like reheating
and all observable effects on the CMB disappear. For $1/3<\wrehbar<1$,
reheating remains unconstrained as for $\wrehbar=0.6$ we still
find $\left \langle \Delta \pi_{\ln\rhoreh}/
\Delta\calP_{\ln\rhoreh}\right \rangle \simeq 1 $ (not
represented). This is in agreement with the expression of the lever arm
in Eq.~\eqref{eq:Rrad}.

\par

Finally, our results show that the Bayesian evidence of a given model
differs for different values of $\wrehbar$, {\ie}, depends on the
assumptions made on reheating. For instance, for loop inflation
$\lip$, the Bayesian evidence varies from $\ln(\calE/\calEbest)\simeq
-0.41$ (inconclusive zone) for $\wrehbar=-0.3$ to $-1.11$ (weakly
disfavored) for $\wrehbar=-0.2$, $-2.59$ for $\wrehbar=0$ (moderately
disfavored) and $-3.27$ for $\wrehbar=0.2$ (moderately
disfavored). This means that in order to estimate the performance of a
model, the details of reheating now matter and must be part of the
model definition.

In conclusion, we have derived the posterior distributions of the
parameters describing the kinematics of the reheating era for nearly
$200$ inflationary scenarios. We have shown that the Planck $2013$ CMB
data put non-trivial constraints on the reheating epoch. The
precise bounds on the reheating parameter, and on the reheating
temperature at fixed equation of state, depend on the model under
consideration. Under the most generic parametrization, we have found
that the Planck data yield to an average reduction of the reheating
prior volume by $40\%$ in the whole space of models tested. 
In more detail, from the results presented here, we can infer the
bounds on $\rhoreh^{1/4}$ for each model of {\EI}. Because of space
limitation, we do not reproduce all of them but it is interesting to
give a few examples. For small field scenarios $\sfi$, a case already
considered in Ref.~\cite{Martin:2010kz}, we find, at $95\%$ of
confidence, $\rhoreh^{1/4} > 400\, \TeV$ for $\wrehbar=-0.3$,
$\rhoreh^{1/4}>90 \, \TeV$ for $\wrehbar=-0.2$, and no constraint for
larger values of $\wrehbar$. Better constraints can be found for other
models. For supergravity brane inflation $\sbi$, one obtains
$\rhoreh^{1/4} > 3.0 \times 10^{6} \, \TeV$ ($\wreh=-0.3$),
$\rhoreh^{1/4} > 1.8 \times 10^{4} \, \TeV$ ($\wreh=-0.2$), and
$\rhoreh^{1/4} > 11\,\GeV$ for $\wreh=0$. For $\wreh=0.6$, the
reheating temperature becomes bounded from above: $\rhoreh^{1/4} < 3.8
\times 10^{11}\,\TeV$ (and, hence, reheating cannot be instantaneous
in that case). Finally, for $\lip$, one obtains upper bounds on the
reheating temperature even for $\wrehbar \le 0$, namely $\rhoreh^{1/4}
< 1.8 \times 10^{7} \TeV$ ($\wrehbar=-0.3$), $\rhoreh^{1/4} < 6.5
\times 10^{7}\, \TeV$ ($\wrehbar=-0.2$), $\rhoreh^{1/4} < 4.0 \times
10^{10}\, \TeV$ ($\wrehbar=0$), and $\rhoreh^{1/4} < 5.1 \times 10^{11}
\, \TeV$ for $\wrehbar=0.2$.

Another result found in this Letter is that the Bayesian evidence of a
model can change in a non negligible way according to the assumptions
made on its reheating properties. This indicates that, with high
accuracy CMB data, reheating details are now important. Obviously,
this will become even more relevant in the case of future CMB
missions~\cite{Martin:2014rqa}. The results presented here represent
the first complete survey of what can be deduced about inflationary
reheating from the Planck data.

\begin{acknowledgments}
The work of C. R. is supported by the ESA Belgian Federal PRODEX Grant
No.~4000103071 and Wallonia-Brussels Federation Grant ARC
No.~11/15-040. The work of V. V. is supported by STFC grant ST/L005573/1.
\end{acknowledgments}

\bibliography{biblio}

\end{document}